\RequirePackage{fixltx2e}
\documentclass[12pt,english]{extarticle}

\usepackage[T1]{fontenc}
\usepackage[latin9]{inputenc}
\usepackage{geometry}
\usepackage{units}
\usepackage{amsmath}
\usepackage{amsthm}
\usepackage{amssymb}
\usepackage{graphicx}

\makeatletter
\theoremstyle{plain}
\newtheorem{thm}{\protect\theoremname}
\theoremstyle{definition}
\newtheorem{defn}[thm]{\protect\definitionname}
\theoremstyle{plain}
\newtheorem{prop}[thm]{\protect\propositionname}
\ifx\proof\undefined
\newenvironment{proof}[1][\protect\proofname]{\par
	\normalfont\topsep6\p@\@plus6\p@\relax
	\trivlist
	\itemindent\parindent
	\item[\hskip\labelsep\scshape #1]\ignorespaces
}{%
	\endtrivlist\@endpefalse
}
\providecommand{\proofname}{Proof}
\fi
\theoremstyle{plain}
\newtheorem{lem}[thm]{\protect\lemmaname}
\theoremstyle{definition}
\newtheorem{problem}[thm]{\protect\problemname}
\theoremstyle{definition}
\newtheorem{example}[thm]{\protect\examplename}
\theoremstyle{plain}
\newtheorem{cor}[thm]{\protect\corollaryname}
\theoremstyle{remark}
\newtheorem{rem}[thm]{\protect\remarkname}

\@ifundefined{date}{}{\date{}}
\makeatother

\usepackage{babel}
\providecommand{\corollaryname}{Corollary}
\providecommand{\definitionname}{Definition}
\providecommand{\examplename}{Example}
\providecommand{\lemmaname}{Lemma}
\providecommand{\problemname}{Problem}
\providecommand{\propositionname}{Proposition}
\providecommand{\remarkname}{Remark}
\providecommand{\theoremname}{Theorem}

\begin{document}
\title{\textbf{\Large{}How to Find New Characteristic-Dependent Linear Rank
Inequalities using Secret Sharing}}
\author{Victor Peña-Macias\thanks{E-mail: vbpenam@unal.edu.co}\\
Departamento de Matemáticas\\
 Universidad Nacional de Colombia\\
Bogotá, Colombia}
\maketitle
\begin{abstract}
Determining information ratios of access structures is an important
problem in secret sharing. Information inequalities and linear rank
inequalities play an important role for proving bounds. Characteristic-dependent
linear rank inequalities are rank inequalities which are true over
vector spaces with specific field characteristic. In this paper, using
ideas of secret sharing, we show a theorem that produces characteristic-dependent
linear rank inequalities. These inequalities can be used for getting
lower bounds on information ratios in linear secret sharing.
\end{abstract}

\section{Introduction}

\emph{Secret sharing} is a cryptographic protocol that consists of
distributing a \emph{secret} in several messages or \emph{shares}
within a group of participants, in such a way that if a group of participants
with access to the secret shares their messages, they can discover
the secret; but if a group of participants that does not have access
to the secret shares their messages, they cannot get any information
about the secret \cite{On the Classification of Ideal Secret Sharing,On Abelian Secret Sharing: duality and separation,Lecture Notes in Secret Sharing,On secret-sharing matroids}.
A specific protocol with this property is called the \emph{secret
sharing scheme}, and the collection of participants with access to
the secret is called the \emph{access structure}. The efficiency of
a scheme is measured by \emph{information ratios} which relates the
size of the secret and the size of the shares. In secret sharing,
it is important to build efficient secret sharing schemes on an access
structure. Therefore, determining the best information ratio, known
as the \emph{optimal information ratio}, is an important task.

A \emph{linear rank inequality} is a linear inequality that is always
satisfied by ranks (dimensions) of subspaces of a vector space over
any field. \emph{Information inequalities} are a sub-class of linear
rank inequalities \cite{Inequalities for Shannon Entropy and Kolmogorov Complexity}.
A \emph{characteristic-dependent linear rank inequality} is like a
linear rank inequality but this is always satisfied by vector spaces
over fields of certain characteristic and does not in general hold
over other characteristics \cite{Achievable Rate Regions for Network Coding,Characteristic-Dependent Linear Rank Inequalities via Complementary Vector,Characteristic-Dependent Linear Rank Inequalities in 21 variables}.
Information inequalities have been useful to estimate lower bounds
on the optimal information ratio of access structures, and linear
rank inequalities have been useful to estimate lower bounds on the
optimal information ratio of access structures in \emph{linear secret
sharing schemes}, i.e. when secret sharing schemes have a linear structure
\cite{Improving the Linear Programming Technique in the Search for Lower Bounds in Secret Sharing}.
To the best of our knowledge, characteristic-dependent linear rank
inequalities have not been used for determining bounds in linear secret
sharing schemes in specific finite fields, but due to the nature of
distinguishing finite fields according to their characteristics, these
inequalities can be useful. One area where these inequalities have
been useful for determing bounds is in \emph{newtork coding} \cite{Lexicographic Products and the Power of non-Linear Network Coding,Achievable Rate Regions for Network Coding,Characteristic Dependent Linear Rank Inequalities and Applications to Network Coding,Characteristic-Dependent Linear Rank Inequalities via Complementary Vector,Linear Programming Problems in Network Coding and Closure Operators}.

\textbf{Contributions. }In \cite{On Abelian Secret Sharing: duality and separation},
Jafari and S. Khazaei developed a technique for proving lower bounds
on access structures in linear secret sharing schemes on finite fields
with a specific characteristic. They present their technique using
access structures or \emph{matroid ports} associated with the \emph{Fano}
and \emph{non-Fano matroids}. We note that this technique can be improved
in order for producing characteristic-dependent linear rank inequalities
that also imply lower bounds on information ratios in linear secret
sharing. For any binary matrix whose determinant is greater than $1$,
we studied some matroid ports associated with the representable matroid
of this matrix. Since the matrix defines different matroids according
to the field where it is defined, we get different access structures;
we establish the properties that depend on the characteristic of the
finite field associated with the matrix. These properties serve as
a guide to define conditions and inequalities that must satisfy vector
spaces of a specific field characteristic. Then, using the vector
deletion technique of Blasiak et al. in \cite{Lexicographic Products and the Power of non-Linear Network Coding},
which was improved in \cite{Characteristic-Dependent Linear Rank Inequalities via Complementary Vector,Characteristic-Dependent Linear Rank Inequalities in 21 variables},
we get a theorem that produces characteristic-dependent linear rank
inequalities. We emphasize that this theorem produces a pair of inequalities
as long as there is a binary matrix whose determinant is greater than
$1$. We also show a class of matrices that satisfy this property
and produce $2\left\lfloor \frac{n-1}{2}\right\rfloor -4$ inequalities
for each $n\geq7$ and we compute some lower bounds of optimal information
ratios associated with matroid ports of these matrices over specific
fields.

\textbf{Organization of the work. }In section 2 and 3, we study concepts
of information theory and secret sharing. In section 4, we show our
method for producing characteristic-dependent linear rank inequalities;
this method is summarized with a theorem. In section 5, some inequalities
are produced and we estimate lower bounds on information ratios of
some access structures in linear secret sharing.

\section{Information Theory}

Let $A_{1}$, $\ldots$, $A_{n}$ be vector subspaces of a finite
dimensional vector space $V$ over a finite field $\mathbb{F}$. Let
$\sum A_{i}$ be the span of $A_{i}$, $i\in I$. We are interested
in tuples of random variables associated to these vector spaces, for
details about the usual construction of these variables, see \cite{Lecture Notes in Secret Sharing,Characteristic-Dependent Linear Rank Inequalities in 21 variables,Inequalities for Shannon Entropy and Kolmogorov Complexity};
we refer as \emph{linear random variables}. There is a correspondence
between entropy of linear random variables and dimension of vector
spaces \cite[Theorem 2]{Inequalities for Shannon Entropy and Kolmogorov Complexity}.
So, we identify the entropy of linear random variables with the dimension
of the associated spaces, i.e. 
\[
\mathrm{H}\left(A_{i}:i\in I\right):=\dim\left(\sum A_{i}\right).
\]
The \emph{mutual information} of $A_{1}$ and $A_{2}$ is given by
\[
\mathrm{I}\left(A_{1};A_{2}\right):=\dim\left(A_{1}\cap A_{2}\right).
\]
The \emph{codimension of $A_{1}$ in $V$} is given by $\mathrm{codim}_{V}\left(A_{1}\right)=\dim\left(V\right)-\dim\left(A_{1}\right)$,
we have 
\[
\mathrm{H}\left(A_{1}\mid A_{2}\right):=\mathrm{codim}_{A_{1}}\left(A_{1}\cap A_{2}\right).
\]
The conditional mutual information is expressed as

\[
\mathrm{I}\left(A_{1};A_{2}\mid A_{3}\right):=\dim\left(A_{1}+A_{3}\right)-\mathrm{codim}_{A_{1}}\left(A_{1}\cap A_{2}\cap A_{3}\right).
\]

The following definition is given to fix ideas about inequalities
and vector spaces.
\begin{defn}
Let $P$ be a proper subset of prime numbers and $I_{1}$, $\ldots$,
$I_{k}$$\subseteq\left[n\right]$. Let $\alpha_{i}\in\mathbb{R}$,
for $1\leq i\leq k$. Consider a linear inequality of the form $\sum\alpha_{i}\mathrm{H}\left(A_{j}:j\in I_{i}\right)\geq0$.
The inequality is called a \emph{characteristic-dependent linear rank
inequality} if it holds for all vector spaces $A_{1}$, $\ldots$,
$A_{n}$ over a finite field whose characteristic is in $P$, and
does not in general hold over other characteristics. Besides, the
inequality is called a \emph{linear rank inequality} if it holds for
all vector spaces.
\end{defn}
The following linear rank inequality is called Ingleton's inequality
\cite{Representation of Matroids}. For any $A_{1}$, $A_{2}$, $A_{3}$,
$A_{4}$ subspaces of a finite dimensional vector space,
\[
\mathrm{I}\left(A_{1};A_{2}\right)\leq\mathrm{I}\left(A_{1};A_{2}\mid A_{3}\right)+\mathrm{I}\left(A_{1};A_{2}\mid A_{4}\right)+\mathrm{I}\left(A_{3};A_{4}\right).
\]

The following inequality \cite{Achievable Rate Regions for Network Coding}
is a characteristic-dependent linear rank inequalities over fields
with characteristic other than $2$:
\[
2\mathrm{H}\left(A_{1}\right)+\mathrm{H}\left(A_{2}\right)+2\mathrm{H}\left(A_{3}\right)\leq\mathrm{H}\left(B_{1}\right)+\mathrm{H}\left(B_{2}\right)+\mathrm{H}\left(B_{3}\right)+\mathrm{H}\left(C\right)
\]
\[
+2\mathrm{H}\left(A_{1}\mid B_{1},C\right)+\mathrm{H}\left(A_{2}\mid B_{2},C\right)+2\mathrm{H}\left(A_{3}\mid A_{1},B_{2}\right)
\]
\[
+3\mathrm{H}\left(B_{2}\mid B_{1},B_{3}\right)+3\mathrm{H}\left(C\mid A_{3},B_{3}\right)+5\mathrm{H}\left(B_{3}\mid A_{1},A_{2}\right)+5\mathrm{H}\left(B_{1}\mid A_{2},A_{3}\right)
\]
\[
+5\left(\mathrm{H}\left(A_{1}\right)+\mathrm{H}\left(A_{2}\right)+\mathrm{H}\left(A_{3}\right)-\mathrm{H}\left(A_{1},A_{2},A_{3}\right)\right).
\]

Let $V=A_{1}\oplus\cdots\oplus A_{n}$ and take a vector subspace
$C$ of $V$ such that 
\[
A_{1}+\cdots+A_{i-1}+C+A_{i+1}+\cdots+A_{n}
\]
is a direct sum for each $i$. We say that $\left(A_{1},\ldots,A_{n},C\right)$
is\emph{ a tuple of complementary vector spaces}. Every vector of
$V$ has a unique representation as a sum of elements of $A_{1}$,
$\ldots$, $A_{n}$. Therefore, $\pi_{I}$ denotes the $I$-projection
function $V\twoheadrightarrow\underset{i\in I}{\bigoplus}A_{i}$ given
by
\[
x=\sum x_{i}\mapsto\underset{i\in I}{\sum}x_{i}.
\]

\begin{prop}
\label{prop: propiedad del subespacio C (de complementarios) es a lo sumo de dim 1}For
any tuple $\left(A_{1},\ldots,A_{n},C\right)$ of complementary vector
spaces, we have 
\[
\mathrm{H}\left(\pi_{I}\left(C\right)\right)=\mathrm{H}\left(C\right)\leq\mathrm{H}\left(A_{i}\right),
\]
for all $i$ and $\emptyset\neq I\subseteq\left[n\right]$.
\end{prop}
\begin{proof}
See proposition 6 in \cite{Characteristic-Dependent Linear Rank Inequalities in 21 variables}.
\end{proof}
\begin{lem}
\label{Capitulo Tecnica 2 Lema elemental para derivar de una estructuras de acceso  unos espacios interesantes}Let
$A_{1}$, $\ldots$, $A_{n}$ and $C$ be vector subspaces of a finite
dimensional vector space $V$, such that
\begin{description}
\item [{-}] $C\leq\sum A_{i}$.
\item [{-}] $C\cap\underset{i\neq k}{\sum}A_{i}=O$ for all $k$.
\end{description}
Then, there exist subspaces $\bar{A}_{i}\leq A_{i}$, for $i=1,\ldots,n$,
such that
\begin{description}
\item [{(i)}] $\bar{A}_{k}\cap\underset{i\neq k}{\sum}A_{i}=O$ for all
$k$.
\item [{(ii)}] $\left(\bar{A}_{1},\ldots,\bar{A}_{n},C\right)$ is a tuple
of complementary vector spaces.
\item [{(iii)}] $\mathrm{H}\left(C\right)=\mathrm{H}\left(\bar{A}_{i}\right)$
for all $i$.
\end{description}
\end{lem}
\begin{proof}
In case $C=O$, we take $\bar{A}_{i}=O$ for all $i$. Otherwise,
we assume $A_{1},\ldots,A_{n},C\neq O$. Let $\left(e_{i}\right)$
be a basis of $C$, we remark that each $e_{i}$ can be written as
$\underset{j}{\sum}e_{i}^{j}$ with $e_{i}^{j}\in A_{i}$ by hypothesis.
Define $\bar{A}_{i}=\left\langle e_{i}^{j}:j\right\rangle $. For
proving (i), we take
\[
x=\underset{i}{\sum}\alpha_{i}e_{k}^{i}\in\bar{A}_{k}\cap\underset{i\neq k}{\sum}A_{i}.
\]
We define $\underset{i}{\sum}\alpha_{i}e_{i}\in C$ for getting
\[
\underset{i}{\sum}\alpha_{i}e_{i}=\underset{i}{\sum}\alpha_{i}\underset{j}{\sum}e_{i}^{j}
\]
\[
\,\,\,\,\,\,\,\,\,\,\,\,\,\,\,\,\,\,\,\,\,\,\,\,\,=\underset{i}{\sum}\alpha_{i}e_{k}^{i}+\underset{j}{\sum}\underset{i\neq k}{\sum}\alpha_{i}e_{i}^{j}
\]
\[
\,\,\,\,\,\,\,\,\,\,\,\,\,\,=x+\underset{j}{\sum}\underset{i\neq k}{\sum}\alpha_{i}e_{i}^{j}.
\]
Thus, $\underset{i}{\sum}\alpha_{i}e_{i}\in C\cap\underset{i\neq k}{\sum}A_{i}$
which implies that $\underset{i}{\sum}\alpha_{i}e_{i}=O$ by hypothesis.
Since $\left(e_{i}\right)$ is a basis, $\alpha_{i}=0$, i.e. $x=O$.
Hence, (i) is true. In particular, this implies that $\bar{A}_{k}\cap\underset{i\neq k}{\sum}\bar{A}_{i}=O$,
and by definition $C\leq\underset{i}{\sum}\bar{A}_{i}$. It follows
that (ii) is true. We also note $\bar{A}_{i}$ is generated by at
most $\mathrm{H}\left(C\right)$-vectors; therefore, by Proposition
\ref{prop: propiedad del subespacio C (de complementarios) es a lo sumo de dim 1}
and (ii), we have (iii) is true. 
\end{proof}
\begin{lem}
\label{Capitulo Tecnica 2 Segundo Lema elemental sobre cotas superiores}For
any vector subspaces $A_{1},\ldots,A_{n}$ of a finite dimensional
vector space $V$, we have
\[
\underset{i}{\sum}\mathrm{H}\left(A_{i}\right)-\mathrm{I}\left(A_{1};\cdots;A_{n}\right)\leq\underset{1<i}{\sum}\mathrm{H}\left(A_{1},A_{i}\right).
\]
\end{lem}
\begin{proof}
The proof is by induction. The case $n=2$ gives a straightforward
information identity. We suppose the case $n-1$ holds, and we show
the case $n$ is true,
\[
\underset{i}{\sum}\mathrm{H}\left(A_{i}\right)-\mathrm{I}\left(A_{1};\cdots;A_{n}\right)=\mathrm{H}\left(A_{n}\right)+\mathrm{H}\left(A_{1}\cap\cdots\cap A_{n-1}\right)
\]
\[
-\mathrm{I}\left(A_{1}\cap\cdots\cap A_{n-1};A_{n}\right)+\underset{i\leq n-1}{\sum}\mathrm{H}\left(A_{i}\right)-\mathrm{I}\left(A_{1};\cdots;A_{n-1}\right)
\]
\[
\,\,\,\,\,\,\,\,\,\,\,\,\,\,\,\,\,\,\leq\mathrm{H}\left(A_{1}\cap\cdots\cap A_{n-1},A_{n}\right)+\underset{1<i\leq n-1}{\sum}\mathrm{H}\left(A_{1},A_{i}\right)\text{ [from cases \ensuremath{n=2} and \ensuremath{n-1}]}
\]
\[
\leq\mathrm{H}\left(A_{1},A_{n}\right)+\underset{1<i\leq n-1}{\sum}\mathrm{H}\left(A_{1},A_{i}\right)
\]
\[
=\underset{1<i}{\sum}\mathrm{H}\left(A_{1},A_{i}\right).
\]
\end{proof}

\section{Secret Sharing}

\emph{Secret Sharing} is an important component in many kinds of cryptographic
protocols \cite{On the Classification of Ideal Secret Sharing,Improving the Linear Programming Technique in the Search for Lower Bounds in Secret Sharing,Lecture Notes in Secret Sharing}.
In a \emph{secret sharing scheme}, a \emph{secret value} is distributed
into \emph{shares} among a set of\emph{ participants} in such a way
that only the \emph{qualified sets} of participants can recover the
secret value.
\begin{defn}
An \emph{access structure}, denoted by $\Gamma$ on a set of participants
$P$, is a monotone increasing family of subsets of $P$. Consider
a special participant $p\notin P$, called dealer. A \emph{secret
sharing scheme} on $P$ with access structure $\Gamma$ is a tuple
of random variables $\Sigma:=\left(S_{x}\right)_{x\in Q}$, where
$Q=P\cup p$, such that the following properties are satisfied:
\begin{description}
\item [{(i)}] $\mathrm{H}\left(S_{p}\right)>0$.
\item [{(ii)}] If $A\in\Gamma$, then $\mathrm{H}\left(S_{p}\mid S_{A}\right)=0$.
\item [{(iii)}] If $A\notin\Gamma$, then $\mathrm{I}\left(S_{p};S_{A}\right)=0$.
\end{description}
\end{defn}
The random variable $S_{p}$ is the \emph{secret value}, and the \emph{shares}
received by the participants are given by the random variables $S_{x}$,
$x\in P$. A set of participants $A$ is said to be \emph{qualified}
if $A\in\Gamma$; and it is said to be \emph{non-qualified} if $A\notin\Gamma$.
A \emph{minimal qualified set} is a qualified set such that any proper
subset is non-qualified. It is clear that an access structure is determined
by the family $\min\mathcal{F}$ of its minimal qualified sets. 
\begin{defn}
The\emph{ information ratio} $\sigma\left(\Sigma\right)$ of the secret
sharing scheme $\Sigma$ is given by 
\[
\sigma\left(\Sigma\right)=\underset{x\in P}{\max}\frac{\mathrm{H}\left(S_{x}\right)}{\mathrm{H}\left(S_{p}\right)}.
\]
The \emph{optimal information ratio} $\sigma\left(\Gamma\right)$
of an access structure $\Gamma$ is the infimum of the information
ratios of all secret sharing schemes for $\Gamma$. The optimal information
ratio using only tuples of linear random variables is denoted by $\lambda\left(\Gamma\right)$.
When we want to specify the characteristic of the field $\mathbb{F}$
or some characteristic field condition, the optimal information ratio
is denoted by $\lambda_{\mathrm{char}\left(\mathbb{F}\right)}\left(\Gamma\right)$.
\end{defn}
We study the following classes of linear programming problems which
are useful for calculating bounds on optimal information ratios \cite{Improving the Linear Programming Technique in the Search for Lower Bounds in Secret Sharing}.
\begin{problem}
\label{prob:problema de programaci=0000F3n lineal para secret sharing}For
any access structure $\Gamma$ on a set $P$ with leader $p\notin P$,
the optimal solution $\kappa\left(\Gamma\right)$ of the linear programming
problem is to calculate $\min\left(v\right)$ such that
\begin{description}
\item [{(i)}] $v\geq f\left(x\right)$ for each $x\in P$.
\item [{(ii)}] $f\left(X\cup p\right)=f\left(X\right)$ for each $X\subseteq P$
with $X\in\Gamma$.
\item [{(iii)}] $f\left(X\cup p\right)=f\left(X\right)+1$ for each $X\subseteq P$
with $X\notin\Gamma$.
\item [{(iv)}] Information inequalities.
\end{description}
\end{problem}
Given a secret sharing scheme $\Sigma=\left(S_{x}\right)_{x\in Q}$,
with access structure $\Gamma$, we consider the mapping given by
$h\left(X\right):=\mathrm{H}\left(S_{X}\right)$, for every $X\subseteq Q$.
We define $f=\frac{1}{h\left(p\right)}h$. The function $f$ satisfies
the conditions of problem 8. Therefore, $f$ is a feasible solution
and we have 
\[
\kappa\left(\Gamma\right)\leq\sigma\left(\Gamma\right).
\]
When we add linear rank inequalities in (iv), we have a linear programming
problem whose optimal solution, denoted by $\kappa^{*}\left(\Gamma\right)$
holds 
\[
\kappa^{*}\left(\Gamma\right)\leq\lambda\left(\Gamma\right).
\]
When we add characteristic-dependent linear rank inequalities the
optimal solution is denoted by $\kappa_{\mathrm{char}\left(\mathbb{F}\right)}^{*}\left(\Gamma\right)$,
we obtain 
\[
\kappa_{\mathrm{char}\left(\mathbb{F}\right)}^{*}\left(\Gamma\right)\leq\lambda_{\mathrm{char}\left(\mathbb{F}\right)}\left(\Gamma\right).
\]
\begin{figure}[h]
\centering{}\includegraphics[scale=0.55]{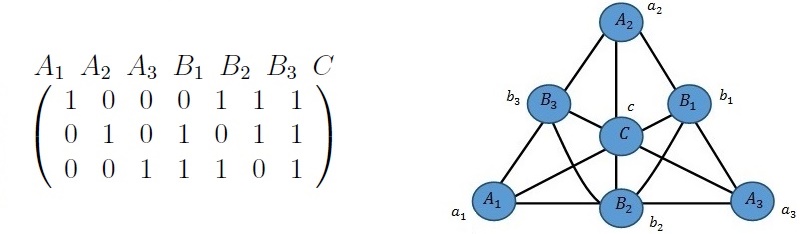}\caption{\label{fig:cap Basics- A-matrix-over  GF(p)  (Fano y non-Fano matrix)}A
matrix over $\mathrm{GF}\left(p\right)$ and \label{capitulo basics port de fano}Fano
matroid.}
\end{figure}

\begin{defn}
A secret sharing scheme $\Sigma=\left(S_{x}\right)_{x\in Q}$ is said
to be\emph{ ideal} if its information ratio is equal to $1$. An access
structure that admits an ideal secret sharing scheme is called \emph{ideal
access structure}.
\end{defn}
Matroids are related to secret sharing, for concepts associated, see
\cite{Matroid Theory}.
\begin{defn}
Given a matroid $\mathcal{M}=\left(Q,r\right)$, where $Q$ is the
ground set and $r$ is the rank function. The \emph{port }of the matroid
$\mathcal{M}$ at $p\in Q$ is the access structure on $P=Q-p$ whose
qualified sets are the sets $X\subseteq P$ satisfying $r\left(X\cup p\right)=r\left(X\right)$.
\end{defn}
The following result connects ideal secret sharing and matroids.
\begin{thm}
\label{Capitulo Basics teorema que relaciona p representacion con secret sharing thm:Let--be}Let
$\Sigma=\left(S_{x}\right)_{x\in Q}$ be an ideal secret sharing scheme
on $P$ with access structure $\Gamma$. Then, the mapping given by
$f\left(X\right)=\nicefrac{\mathrm{H}\left(S_{X}\right)}{\mathrm{H}\left(S_{p}\right)}$
for each $X\subseteq Q$ is the rank function of a matroid $\mathcal{M}$
with ground set $Q$. Moreover, $\Gamma$ is the port of the matroid
$\mathcal{M}$ at $p$.
\end{thm}
As a consequence, every ideal access structure is a matroid port.
A known result about $\kappa$ and matroid ports is as follows.
\begin{thm}
Let $\Gamma$ be an access structure. Then, $\Gamma$ is a matroid
port if and only if $\kappa\left(\Gamma\right)=1$. Moreover, $\kappa\left(\Gamma\right)\geq\frac{3}{2}$
if $\Gamma$ is not a matroid port.
\end{thm}
\begin{example}
\label{capitulo basics port del fano matroid}The port of the Fano
matroid at $c$, according figure \ref{capitulo basics port de fano},
is given by the minimum qualified sets 
\[
\min\mathcal{F}:=\left\{ a_{1}b_{1},a_{2}b_{2},a_{3}b_{3},a_{1}a_{2}a_{3},a_{1}b_{2}b_{3},b_{1}a_{2}b_{3},b_{1}b_{2}a_{3}\right\} .
\]
The columns of a matrix of representation of Fano matroid define an
ideal linear secret sharing scheme over fields whose characteristic
is two, and therefore the ports of the Fano matroid are ideal. We
have:
\begin{itemize}
\item $\sigma\left(\mathcal{F}\right)=\lambda\left(\mathcal{F}\right)=\lambda_{\mathrm{char}\left(\mathbb{F}\right)=2}\left(\mathcal{F}\right)=1,$
\item $\kappa\left(\Gamma\right)=\kappa^{*}\left(\Gamma\right)=\kappa_{\mathrm{char}\left(\mathbb{F}\right)=2}^{*}\left(\Gamma\right)=1$.
\end{itemize}
It is more hard for showing \cite{On Abelian Secret Sharing: duality and separation,Decomposition Constructions for Secret}:
\begin{itemize}
\item $\lambda_{\mathrm{char}\left(\mathbb{F}\right)\neq2}\left(\mathcal{F}\right)=\kappa_{\mathrm{char}\left(\mathbb{F}\right)\neq2}^{*}\left(\Gamma\right)=\frac{4}{3}.$
\end{itemize}
\end{example}

\section{Producing inequalities}

Consider any $n\times n$ binary matrix $B=\left(b_{ji}\right)$,
we write the $i$-th column as $e_{S_{i}}$ where 
\[
S_{i}=\left\{ j:b_{ji}=1\right\} .
\]
We then define the sets:
\[
\mathcal{B}':=\left\{ e_{S_{i}}:1<\left|S_{i}\right|<n\right\} ,
\]
\[
\mathcal{B}'':=\left\{ e_{S_{i}}:\left|S_{i}\right|=1\right\} ,
\]
\[
\mathcal{B}''':=\left\{ e_{S_{i}}:\left|S_{i}\right|=n\right\} .
\]
In the following, we suppose that $\left|\det\left(B\right)\right|=t>1$,
for some $t\in\mathbb{N}$ and $\mathcal{B}'''$ is empty.

These matrices can be used to define matroid ports which are ideal
over some fields; they are representable matroids.

We consider $n+\left|\mathcal{B}'\right|$ participants labeled as
follows
\[
P:=\left\{ a_{e_{i}}:i\in\left[n\right]\right\} \cup\left\{ b_{e_{S_{j}}}:e_{S_{j}}\in\mathcal{B}'\right\} .
\]
We remark that $n$ participants are labeled using the canonical basis
$\left(e_{i}\right)$ and $\left|\mathcal{B}'\right|$ participants
are labeled using the columns of $B$ in $\mathcal{B}'$.

Now, consider any access structure on $n+\left|\mathcal{B}'\right|$
participants such that:
\begin{itemize}
\item The following set is a subclass of the collection of minimal qualified
sets,
\[
\left\{ \left(a_{e_{i}}\right)_{i\notin S_{j}}b_{e_{S_{j}}}:e_{S_{j}}\in\mathcal{B}'\right\} \cup\left\{ a_{e_{1}}\cdots a_{e_{n}}\right\} .
\]
\item The following set is a subclass of the class of non-qualified sets,
\[
\left\{ \left(a_{e_{i}}\right)_{i\in S_{j}}b_{e_{S_{j}}}:e_{S_{j}}\in\mathcal{B}'\right\} .
\]
\end{itemize}
Let $P_{B}$ be the subset of participants labeled by the columns
of $B$. We produce two different classes of access structures according
to add $P_{B}$ to the subclass of minimal qualified sets or to the
subclass of non-qualified sets. There are several access structures
with these properties, in the next section we show an example.

The classes of access structures defined above can be used as a guide
for determining properties or conditions that must be satisfied by
associated vector spaces in order to derive some inequalities. Each
vector space can be thought of as follows:
\begin{itemize}
\item $A_{e_{i}}$ is associated to $a_{e_{i}}$.
\item $B_{e_{i}}$ is associated to $b_{e_{i}}$ .
\item $C$ is associated to the dealer $c\notin P$.
\end{itemize}
\begin{prop}
\label{prop:conversi=0000F3n de conjuntos minimales calificados}Let
$A_{e_{i}}$, for $i\in\left[n\right]$, $B_{e_{S_{j}}}$, for $e_{S_{j}}\in\mathcal{B}'$,
and $C$ be vector subspaces of a vector space $V$ such that
\begin{description}
\item [{-}] $C\leq A_{\left[e_{n}\right]}\cap\left(\underset{i\notin S_{j}}{\sum}A_{e_{i}}+B_{e_{S_{j}}}\right)$
for each $j$.
\item [{-}] $C\cap A_{\left[e_{n}\right]-e_{i}}=O$ for each $i$.
\item [{-}] $C\cap\left(\underset{i\notin S_{j},i\neq k}{\sum}A_{e_{i}}+B_{e_{S_{j}}}\right)=O$
for each $e_{S_{j}}\in\mathcal{B}$ and $k\notin S_{j}$.
\end{description}
Then, we have vector subspaces $\bar{A}_{e_{i}}\leq A_{e_{i}}$, $i\in\left[n\right]$;
$A_{e_{i}}^{S_{j}}\leq A_{e_{i}}$ and $\hat{B}_{e_{S_{j}}}\leq B_{e_{S_{j}}}$
for each $e_{S_{j}}\in\mathcal{B}'$ and $i\notin S_{j}$ such that
\begin{description}
\item [{-}] $\left(\bar{A}_{e_{1}},\cdots,\bar{A}_{e_{n}},C\right)$ is
a tuple of complementary vector spaces.
\item [{-}] $\left(A_{e_{i}}^{S_{j}},\hat{B}_{e_{S_{j}}},C:i\notin S_{j}\right)$,
for each $e_{S_{j}}\in\mathcal{B}'$, is a tuple of complementary
vector spaces.
\item [{-}] The dimension of any of these subspaces is $\mathrm{H}\left(C\right)$.
\item [{-}] These subspaces are unique except isomorphisms.
\end{description}
\end{prop}
This proposition is a consequence of Lemma \ref{Capitulo Tecnica 2 Lema elemental para derivar de una estructuras de acceso  unos espacios interesantes}
and expresses the fact that in the access structure the participants
$a_{e_{1}}\cdots a_{e_{n}}$, $\left(a_{e_{i}}\right)_{i\notin S_{j}}b_{e_{S_{j}}}$,
for all $e_{S_{j}}\in\mathcal{B}'$, are minimal qualified sets.

The following linear mapping is well-defined under hypothesis of previous
proposition:
\[
\begin{array}{c}
\varphi_{B}:C\rightarrow\underset{i}{\bigoplus}\frac{\bar{A}_{e_{i}}}{\left(\bigcap\mathcal{A}_{e_{i}}\right)\cap\bar{A}_{e_{i}}}\\
c\longmapsto\varphi_{B}\left(c\right):=\underset{i}{\sum}\left[a_{i}\right]_{\left(\bigcap\mathcal{A}_{e_{i}}\right)\cap\bar{A}_{e_{i}}}
\end{array},
\]
where $c=\underset{i}{\sum}a_{i}$ with $a_{i}\in\bar{A}_{e_{i}}$
and $\mathcal{A}_{e_{i}}:=\left\{ A_{e_{i}}^{S_{j}}:i\notin S_{j}\text{ for some }j\right\} $;
we take $\mathcal{A}_{e_{i}}$ as $\left\{ O\right\} $, in case that
$i\in S_{j}$ for all $j$. We remark that there is a correspondence
between $\mathcal{A}_{e_{i}}$ and the subset of columns of $B$ given
by $\mathcal{B}_{e_{i}}:=\left\{ e_{S_{j}}:i\notin S_{j}\right\} $;
we take $\mathcal{B}_{e_{i}}$ as $\left\{ O\right\} $ in case that
$i\in S_{j}$ for all $j$.
\begin{lem}
\label{lema desigualdad con kerner}For any vector subspaces $A_{e_{1}},\ldots,A_{e_{n}}$,
$B_{e_{S_{j_{1}}}},\ldots,B_{e_{S_{j_{\left|\mathcal{B}'\right|}}}}$
and $C$ of a finite dimensional vector space $V$ such that
\begin{description}
\item [{(i)}] $C\leq A_{\left[e_{n}\right]}\cap\left(\underset{i\notin S_{j}}{\sum}A_{e_{i}}+B_{e_{S_{j}}}\right)$
for each $e_{S_{j}}\in\mathcal{B}'$.
\item [{(ii)}] $C\cap A_{\left[e_{n}\right]-e_{i}}=O$ for each $i$.
\item [{(iii)}] $C\cap\left(\underset{i\notin S_{j},i\neq k}{\sum}A_{e_{i}}+B_{e_{S_{j}}}\right)=O$
for each $e_{S_{j}}\in\mathcal{B}$ and $k\notin S_{j}$.
\end{description}
Then,
\[
\left[1+\underset{i}{\sum}\left|\mathcal{B}_{e_{i}}\right|\right]\mathrm{H}\left(C\right)\leq\underset{i}{\sum}\left|\mathcal{B}_{e_{i}}\right|\mathrm{H}\left(A_{e_{i}}\right)+\mathrm{H}\left(\mathrm{ker}\left(\varphi_{B}\right)\right).
\]
\end{lem}
\begin{proof}
From mapping $\varphi_{B}$, we can derive the inequality
\[
\mathrm{H}\left(\frac{C}{\mathrm{ker}\left(\varphi_{B}\right)}\right)\leq\underset{i}{\sum}\mathrm{H}\left(\frac{\bar{A}_{e_{i}}}{\left(\bigcap\mathcal{A}_{e_{i}}\right)\cap\bar{A}_{e_{i}}}\right).
\]
So
\[
\mathrm{H}\left(C\right)-\mathrm{H}\left(\mathrm{ker}\left(\varphi_{B}\right)\right)\leq\underset{i}{\sum}\left[\mathrm{H}\left(\bar{A}_{e_{i}}\right)-\mathrm{I}\left(\bar{A}_{e_{i}};\bigcap\mathcal{A}_{e_{i}}\right)\right].
\]
Then
\[
\mathrm{H}\left(C\right)-\mathrm{H}\left(\mathrm{ker}\varphi_{B}\right)+\underset{i}{\sum}\underset{e_{S_{j}}\in\mathcal{B}_{e_{i}}}{\sum}\mathrm{H}\left(A_{e_{i}}^{S_{j}}\right)\leq\underset{i}{\sum}\left[\mathrm{H}\left(\bar{A}_{e_{i}}\right)+\underset{e_{S_{j}}\in\mathcal{B}_{e_{i}}}{\sum}\mathrm{H}\left(A_{e_{i}}^{S_{j}}\right)-\mathrm{I}\left(\bar{A}_{e_{i}};\bigcap\mathcal{A}_{e_{i}}\right)\right],
\]
\[
\leq\underset{i}{\sum}\underset{e_{S_{j}}\in\mathcal{B}_{e_{i}}}{\sum}\mathrm{H}\left(\bar{A}_{e_{i}},A_{e_{i}}^{S_{j}}\right),\,\,\text{[from Lemma \ref{Capitulo Tecnica 2 Segundo Lema elemental sobre cotas superiores}]}.
\]
Since $\mathrm{H}\left(A_{e_{i}}^{S_{j}}\right)=\mathrm{H}\left(C\right)$,
$\underset{e_{S_{j}}\in\mathcal{B}_{e_{i}}}{\sum}1=\left|\mathcal{B}_{e_{i}}\right|$
and $\bar{A}_{e_{i}},A_{e_{i}}^{S_{j}}\leq A_{e_{i}}$, we get
\[
\mathrm{H}\left(C\right)-\mathrm{H}\left(\mathrm{ker}\left(\varphi_{B}\right)\right)+\underset{i}{\sum}\left|\mathcal{B}_{e_{i}}\right|\mathrm{H}\left(C\right)\leq\underset{i}{\sum}\left|\mathcal{B}_{e_{i}}\right|\mathrm{H}\left(A_{e_{i}}\right),
\]
which implies the desired inequality.
\end{proof}
\begin{lem}
\label{Capitulo Tecnica 2 Lema importante que usa estructura de acceso de no divisores de t}For
any vector subspaces $A_{e_{1}},\ldots,A_{e_{n}}$, $B_{e_{S_{j_{1}}}},\ldots,B_{e_{S_{j_{\left|\mathcal{B}'\right|}}}}$
and $C$ of a finite dimensional vector space $V$ over a finite field
$\mathbb{F}$ whose characteristic does not divide $t$, such that
\begin{description}
\item [{(i)}] $C\leq A_{\left[e_{n}\right]}\cap\left(\underset{i\notin S_{j}}{\sum}A_{e_{i}}+B_{e_{S_{j}}}\right)$,
for each $e_{S_{j}}\in\mathcal{B}'$.
\item [{(ii)}] $C\cap A_{\left[e_{n}\right]-e_{i}}=C\cap\left(\underset{i\in S_{j}}{\sum}A_{e_{i}}+B_{e_{S_{j}}}\right)=C\cap\left(\underset{i\notin S_{j},i\neq k}{\sum}A_{e_{i}}+B_{e_{S_{j}}}\right)=O$,
for all $e_{S_{j}}\in\mathcal{B}'$ and $k\notin S_{j}$ .
\item [{(iii)}] $C\cap\mathcal{B}=O$, where $\mathcal{B}$ is the sum
of all vector subspaces indexed by the columns of $B$.
\end{description}
Then
\[
\mathrm{ker}\left(\varphi_{B}\right)=O.
\]
\end{lem}
\begin{proof}
We take $c=\underset{i}{\sum}a_{i}\in C$ such that $\varphi_{B}\left(c\right)=O$,
where $a_{i}\in\bar{A}_{i}$. We have to show $c=O$. By definition
of $\varphi_{B}$, 
\[
a_{i}\in\bar{A}_{e_{i}}\cap\left(\underset{e_{S_{j}}\in\mathcal{B}_{e_{i}}}{\bigcap}A_{e_{i}}^{S_{j}}\right).
\]
Hence, $a_{i}\in A_{e_{i}}^{S_{j}}$ for all $e_{S_{j}}\in\mathcal{B}_{e_{i}}$,
and therefore 
\[
\underset{i\notin S_{j}}{\sum}a_{i}\in\underset{i\notin S_{j}}{\sum}A_{e_{i}}^{S_{j}}.
\]
From (i) in Lemma \ref{Capitulo Tecnica 2 Lema elemental para derivar de una estructuras de acceso  unos espacios interesantes},
there exists $b_{j}\in\hat{B}_{e_{S_{j}}}$ for each $e_{S_{j}}\in\mathcal{B}'$
such that $\underset{i\notin S_{j}}{\sum}a_{i}+b_{j}\in C$. Hence,
\[
\underset{i\in S_{j}}{\sum}a_{i}-b_{j}=\underset{i}{\sum}a_{i}-\left(\underset{i\notin S_{j}}{\sum}a_{i}+b_{j}\right)\in C\cap\left(\underset{i\in S_{j}}{\sum}A_{e_{i}}+B_{e_{S_{j}}}\right).
\]
From (ii), this implies
\[
\underset{i\in S_{j}}{\sum}a_{i}=b_{j}\text{ for all \ensuremath{e_{S_{j}}\in\mathcal{B}'}}.
\]
These equalities define the following linear system of equations
\begin{equation}
B^{T}\left(\begin{array}{c}
a_{1}\\
\vdots\\
a_{n}
\end{array}\right)=\left(\begin{array}{c}
b_{1}\\
\vdots\\
b_{\left|\mathcal{B}'\right|}\\
b^{1}\\
\vdots\\
b^{\left|\mathcal{B}''\right|}
\end{array}\right),\label{eq:ecuaci=0000F3n central de la tecnica 2}
\end{equation}
where the vectors $b^{1},\ldots,b^{\left|\mathcal{B}''\right|}$ are
omitted when $\mathcal{B}''$ is empty; in other case, $b^{i}:=a_{i}$,
for $e_{i}\in\mathcal{B}''$. Since $\mathrm{char}\left(\mathbb{F}\right)$
does not divide $t=\left|\det\left(B\right)\right|$, the matrix $B^{T}$
is non-singular. Therefore, each $a_{i}$ can be written as a linear
combination of $b_{1}$, $\ldots$, $b_{\left|\mathcal{B}'\right|}$,
$b^{1}$, $\ldots$, $b^{\left|\mathcal{B}''\right|}$, which implies
that $c\in\mathcal{B}$. From (iii), we get $c=O$.
\end{proof}
\begin{cor}
\label{Capitulo Tecnica 2 Corolario  importante que usa estructura de acceso de divisores de t}For
any vector subspaces $A_{e_{1}},\ldots,A_{e_{n}}$, $B_{e_{S_{j_{1}}}},\ldots,B_{e_{S_{j_{\left|\mathcal{B}'\right|}}}}$
and $C$ of a finite dimensional vector space $V$ over a finite field
$\mathbb{F}$, such that
\begin{description}
\item [{(i)}] $C\leq A_{\left[e_{n}\right]}\cap\left(\underset{i\notin S_{j}}{\sum}A_{e_{i}}+B_{e_{S_{j}}}\right)$,
for all $e_{S_{j}}\in\mathcal{B}'$.
\item [{(ii)}] $C\cap A_{\left[e_{n}\right]-e_{i}}=C\cap\left(\underset{i\in S_{j}}{\sum}A_{e_{i}}+B_{e_{S_{j}}}\right)=C\cap\left(\underset{i\notin S_{j},i\neq k}{\sum}A_{e_{i}}+B_{e_{S_{j}}}\right)=O$,
for all $e_{S_{j}}\in\mathcal{B}'$ and $k\notin S_{j}$.
\end{description}
Then, the mapping
\[
\begin{array}{c}
\phi_{B}^{k}:\mathrm{ker}\left(\varphi_{B}\right)\rightarrow B_{e_{S_{k}}}\\
c\longmapsto\phi_{B}^{k}\left(c\right):=\underset{i\in S_{k}}{\sum}a_{i}=b_{k}
\end{array}
\]
is an one-to-one well-defined linear function for each $e_{S_{k}}\in\mathcal{B}'$.
Also, if the $k$-th column of $B$ is a linear combination of the
columns of the submatrix of $B$ denoted by $B_{X}$, $k\notin X$.
Then,
\[
\phi_{B}^{k}\left(\mathrm{ker}\left(\varphi_{B}\right)\right)\subseteq\underset{e_{i}\in B_{X}\cap\mathcal{B}''}{\sum}A_{e_{i}}+\underset{e_{S_{i}}\in B_{X}\cap\mathcal{B}'}{\sum}B_{e_{S_{i}}}.
\]
\end{cor}
\begin{proof}
We can follow line-by-line the proof of the previous lemma to obtain
that there exists a unique $b_{k}\in\left(\underset{i\in S_{k}}{\sum}\bar{A}_{i}\right)\cap\hat{B}_{e_{S_{k}}}\subseteq B_{e_{S_{k}}}$.
So $\phi_{B}^{k}$ is well-defined. Since the written of each $c\in\mathrm{ker}\left(\varphi_{B}\right)$
is unique, $\phi_{B}^{k}$ is also an one-to-one linear mapping. Also,
if the $k$-th column of $B$ is a linear combination of the columns
of $B_{X}$, from equation (\ref{eq:ecuaci=0000F3n central de la tecnica 2}),
we have that $b_{k}$ is a linear combination of $\left(b_{i}\right)_{i\in B_{X}}\cup\left(b^{i}\right)_{i\in B_{X}}$
. Therefore, $b_{k}\in\underset{e_{i}\in B_{X}\cap\mathcal{B}''}{\sum}A_{e_{i}}+\underset{e_{S_{i}}\in B_{X}\cap\mathcal{B}'}{\sum}B_{e_{S_{i}}}$.
\end{proof}
We finally show a theorem that can produce characteristic-dependent
linear rank inequalities as long as there are suitable binary matrices.
\begin{thm}
\label{thm:Teorema central del metodo 2}For a $n\times n$ binary
matrix $B$ such that $\mathcal{B}'''=\emptyset$ and $\left|\det\left(B\right)\right|=t\in\mathbb{N}$,
$t>1$. Let $A_{e_{1}},\ldots,A_{e_{n}}$, $B_{e_{S_{j_{1}}}},\ldots,B_{e_{S_{j_{\left|\mathcal{B}'\right|}}}}$
and $C$ be vector subspaces of a finite dimensional vector space
$V$ over $\mathbb{F}.$ We have
\begin{description}
\item [{-}] The following inequality is a characteristic-dependent linear
rank inequality over fields whose characteristic does not divide $t$,
\[
\mathrm{H}\left(C\right)\leq\frac{1}{1+\underset{i}{\sum}\left|\mathcal{B}_{e_{i}}\right|}\underset{i}{\sum}\left|\mathcal{B}_{e_{i}}\right|\mathrm{H}\left(A_{e_{i}}\right)+\mathrm{H}\left(C\mid A_{\left[e_{n}\right]}\right)+\mathrm{I}\left(C;A_{e_{i}},B_{e_{S_{j}}}:e_{i}\in\mathcal{B}'',e_{S_{j}}\in\mathcal{B}'\right)
\]
\[
+\underset{i}{\sum}\mathrm{I}\left(C;A_{\left[e_{n}\right]-e_{i}}\right)+\underset{e_{S_{h}}\in\mathcal{B}',i\notin S_{h}.}{\sum}\mathrm{I}\left(C;A_{e_{j}},B_{e_{S_{h}}}:j\notin S_{h},j\neq i\right)
\]
\[
+\underset{e_{S_{j}}\in\mathcal{B}'}{\sum}\left[\mathrm{H}\left(C\mid A_{e_{i}},B_{e_{S_{j}}},i\notin S_{j}\right)+\mathrm{I}\left(C;A_{e_{i}},B_{e_{S_{j}}}:i\in S_{j}\right)\right].
\]
\item [{-}] Fixed $k\in\left[n\right]$ such that $e_{S_{k}}\in\mathcal{B}'$.
The following inequality is a characteristic-dependent linear rank
inequality over fields whose characteristic divides $t$,
\[
\mathrm{H}\left(C\right)\leq\frac{1}{2+\underset{i}{\sum}\left|\mathcal{B}_{e_{i}}\right|}\left[\underset{i}{\sum}\left|\mathcal{B}_{e_{i}}\right|\mathrm{H}\left(A_{e_{i}}\right)+\mathrm{H}\left(B_{e_{S_{k}}}\right)\right]+\mathrm{H}\left(C\mid A_{e_{i}},B_{e_{S_{j}}}:e_{i}\in\mathcal{B}'',e_{S_{j}}\in\mathcal{B}'\right)
\]
\[
+\mathrm{H}\left(C\mid A_{\left[e_{n}\right]}\right)+\underset{e_{S_{i}}\in\mathcal{B}'}{\sum}\mathrm{H}\left(C\mid A_{e_{j}},B_{e_{S_{i}}}:j\notin S_{i}\right)
\]
\[
+\underset{i}{\sum}\mathrm{I}\left(C;A_{\left[e_{n}\right]-e_{i}}\right)+\underset{e_{S_{h}}\in\mathcal{B}',i\notin S_{h}.}{\sum}\mathrm{I}\left(C;A_{e_{j}},B_{e_{S_{h}}}:j\notin S_{h},j\neq i\right)
\]
\[
+\underset{e_{S_{j}}\in\mathcal{B}'}{\sum}\mathrm{I}\left(C;A_{e_{i}},B_{e_{S_{j}}}:i\in S_{j}\right)+\underset{i}{\sum}\mathrm{I}\left(C;A_{e_{j}},B_{e_{S_{i}}}:e_{S_{j}}\in\mathcal{B}',e_{j}\in\mathcal{B}'',j\neq i\right).
\]
\end{description}
\end{thm}
The inequalities do not in general hold over fields whose characteristic
is different to the mentioned. Counter examples would be in $V=\text{GF}\left(p\right)^{n}$,
take the vector spaces $A_{e_{i}}=\left\langle e_{i}\right\rangle $,
$e_{i}\in\left[e_{n}\right]$, $B_{e_{S_{j}}}=\left\langle e_{S_{j}}\right\rangle $,
$e_{S_{j}}\in\mathcal{B}'$, and $C=\left\langle \sum e_{i}\right\rangle $
Then, when $p$ divides $t$, the first inequality does not hold;
and when $p$ does not divide $t$, the second inequality does not
hold.
\begin{proof}
To prove the first inequality: let $\mathbb{F}$ be a finite field
whose characteristic does not divide $t$. Let
\[
C^{\left\langle 0\right\rangle }:=C\cap A_{\left[e_{n}\right]}\cap\left[\underset{e_{S_{j}}\in\mathcal{B}'}{\bigcap}\left(\underset{i\notin S_{j}}{\sum}A_{e_{i}}+B_{e_{S_{j}}}\right)\right].
\]
We have 
\[
\mathrm{H}\left(C\mid C^{\left\langle 0\right\rangle }\right)\leq\mathrm{H}\left(C\mid A_{\left[e_{n}\right]}\right)+\underset{e_{S_{j}}\in\mathcal{B}'}{\sum}\mathrm{H}\left(C\mid A_{e_{i}},B_{e_{S_{j}}}:i\notin S_{j}\right).
\]
Recursively, for $i\in\left[n\right]$, denote by $C^{\left\langle i\right\rangle }$,
a subspace of $C^{\left\langle i-1\right\rangle }$ which is a complementary
space to $\underset{j\neq i}{\sum}A_{e_{j}}$ in 
\[
C^{\left\langle i-1\right\rangle }+\underset{j\neq i}{\sum}A_{e_{j}}.
\]
We have 
\[
\mathrm{H}\left(C^{\left\langle i-1\right\rangle }\mid C^{\left\langle i\right\rangle }\right)\leq\mathrm{I}\left(C;A_{e_{j}}:j\neq i\right).
\]
Let $C_{e_{S_{j_{1}}}}^{\left[0\right]}:=C^{\left\langle n\right\rangle }$
and recursively, for each $i\notin S_{j_{1}}$, we denote by $C_{e_{S_{j_{1}}}}^{\left[i\right]}$,
a subspace of $C_{e_{S_{j_{1}}}}^{\left[i-1\right]}$ which is a complementary
space to $\underset{j\notin S_{j_{1}},j\neq i}{\sum}A_{e_{j}}+B_{e_{S_{j_{1}}}}$
in 
\[
C_{e_{S_{j_{1}}}}^{\left[i-1\right]}+\underset{j\notin S_{j_{1}},j\neq i}{\sum}A_{e_{j}}+B_{e_{S_{j_{1}}}}.
\]
We have 
\[
\mathrm{H}\left(C_{e_{S_{j_{1}}}}^{\left[i-1\right]}\mid C_{e_{S_{j_{1}}}}^{\left[i\right]}\right)\leq\mathrm{I}\left(C;A_{e_{j}},B_{e_{S_{j_{1}}}}:j\notin S_{j_{1}},j\neq i\right).
\]
In a similar way, we define $C_{e_{S_{j_{2}}}}^{\left[0\right]}=C_{e_{S_{j_{1}}}}^{\left[n\right]}$,
$\ldots$, $C_{e_{S_{j_{\left|\mathcal{B}'\right|}}}}^{\left[0\right]}=C_{e_{S_{j_{\left|\mathcal{B}'\right|-1}}}}^{\left[n\right]}$
until to find a subspace $C^{\left(0\right)}:=C_{e_{S_{j_{\left|\mathcal{B}'\right|}}}}^{\left[n\right]}$
that holds
\[
\mathrm{H}\left(C^{\left\langle n\right\rangle }\mid C^{\left(0\right)}\right)\leq\underset{e_{S_{h}}\in\mathcal{B}',i\notin S_{h}.}{\sum}\mathrm{I}\left(C;A_{e_{j}},B_{e_{S_{h}}}:j\notin S_{h},j\neq i\right).
\]
Recursively, for $i$, with $e_{S_{i}}\in\mathcal{B}'$, we denote
by $C^{\left(i\right)}$, a subspace of $C^{\left(i-1\right)}$ which
is a complementary space to $\underset{j\in S_{i}}{\sum}A_{e_{j}}+B_{e_{S_{i}}}$
in 
\[
C^{\left(i-1\right)}+\left(\underset{j\in S_{i}}{\sum}A_{e_{j}}+B_{e_{S_{i}}}\right).
\]
We also have
\[
\mathrm{H}\left(C^{\left(i-1\right)}\mid C^{\left(i\right)}\right)\leq\mathrm{I}\left(C^{\left(i-1\right)};A_{e_{j}},B_{e_{S_{i}}}:j\in S_{i}\right).
\]
Define by $\hat{C}$, a subspace of $C^{\left(\left|\mathcal{B}'\right|\right)}$
which is a complementary space to
\[
\mathcal{B}=\left(\underset{e_{S_{i}}\in\mathcal{B}'}{\sum}B_{e_{S_{i}}}\right)+\left(\underset{e_{i}\in\mathcal{B}''}{\sum}A_{e_{i}}\right)
\]
in $C^{\left(\left|\mathcal{B}'\right|\right)}+\mathcal{B}$. We have
\[
\mathrm{H}\left(C^{\left(\left|\mathcal{B}'\right|\right)}\mid\hat{C}\right)\leq\mathrm{I}\left(C;A_{e_{i}},B_{e_{S_{j}}}:e_{i}\in\mathcal{B}'',e_{S_{j}}\in\mathcal{B}'\right).
\]
Hence, 
\[
\mathrm{H}\left(C\mid\hat{C}\right)=\mathrm{H}\left(C\mid C^{\left\langle 0\right\rangle }\right)+\mathrm{H}\left(C^{\left\langle 0\right\rangle }\mid C^{\left\langle n\right\rangle }\right)+\mathrm{H}\left(C^{\left\langle n\right\rangle }\mid C^{\left(0\right)}\right)
\]
\[
+\mathrm{H}\left(C^{\left(0\right)}\mid C^{\left(\left|\mathcal{B}'\right|\right)}\right)+\mathrm{H}\left(C^{\left(\left|\mathcal{B}'\right|\right)}\mid\hat{C}\right)
\]
\[
\leq\mathrm{H}\left(C\mid A_{\left[e_{n}\right]}\right)+\underset{e_{S_{j}}\in\mathcal{B}'}{\sum}\mathrm{H}\left(C\mid A_{e_{i}},B_{e_{S_{j}}}:i\notin S_{j}\right)
\]
\[
+\underset{i}{\sum}\mathrm{I}\left(C;A_{\left[e_{n}\right]-e_{i}}\right)+\underset{e_{S_{h}}\in\mathcal{B}',i\notin S_{h}.}{\sum}\mathrm{I}\left(C;A_{e_{j}},B_{e_{S_{h}}}:j\notin S_{h},j\neq i\right)
\]
\[
+\underset{e_{S_{j}}\in\mathcal{B}'}{\sum}\mathrm{I}\left(C;A_{e_{i}},B_{e_{S_{j}}}:i\in S_{j}\right)+\mathrm{I}\left(C;A_{e_{i}},B_{e_{S_{j}}}:e_{i}\in\mathcal{B}'',e_{S_{j}}\in\mathcal{B}'\right).
\]
Since $A_{e_{1}},\ldots,A_{e_{n}}$, $B_{e_{S_{j_{1}}}},\ldots,B_{e_{S_{j_{\left|\mathcal{B}'\right|}}}}$
and $\hat{C}$ satisfy hypothesis in Lemma \ref{Capitulo Tecnica 2 Lema importante que usa estructura de acceso de no divisores de t},
we have $\mathrm{ker}\left(\varphi_{B}\right)=O$. Therefore, as these
spaces also satisfy hypothesis in Lemma \ref{lema desigualdad con kerner},
it follows
\[
\left[1+\underset{i}{\sum}\left|\mathcal{B}_{e_{i}}\right|\right]\mathrm{H}\left(\hat{C}\right)\leq\underset{i}{\sum}\left|\mathcal{B}_{e_{i}}\right|\mathrm{H}\left(A_{e_{i}}\right).
\]
Using the last two inequalities, we can obtain the described inequality:
\[
\mathrm{H}\left(C\right)-\mathrm{H}\left(C\mid A_{\left[e_{n}\right]}\right)-\underset{e_{S_{j}}\in\mathcal{B}'}{\sum}\mathrm{H}\left(C\mid A_{e_{i}},B_{e_{S_{j}}}:i\notin S_{j}\right)
\]
\[
-\underset{i\in\left[n\right]}{\sum}\mathrm{I}\left(C;A_{\left[e_{n}\right]-e_{i}}\right)-\underset{e_{S_{h}}\in\mathcal{B}',i\notin S_{h}.}{\sum}\mathrm{I}\left(C;A_{e_{j}},B_{e_{S_{h}}}:j\notin S_{h},j\neq i\right)
\]
\[
-\underset{e_{S_{j}}\in\mathcal{B}'}{\sum}\mathrm{I}\left(C;A_{e_{i}},B_{e_{S_{j}}}:i\in S_{j}\right)-\mathrm{I}\left(C;A_{e_{i}},B_{e_{S_{j}}}:e_{i}\in\mathcal{B}'',e_{S_{j}}\in\mathcal{B}'\right)\leq\mathrm{H}\left(\hat{C}\right)
\]
\[
\leq\frac{1}{1+\underset{i}{\sum}\left|\mathcal{B}_{e_{i}}\right|}\underset{i}{\sum}\left|\mathcal{B}_{e_{i}}\right|\mathrm{H}\left(A_{e_{i}}\right).
\]
To prove the second inequality, let $k\in\left[n\right]$ such that
$e_{S_{k}}\in\mathcal{B}'$ and let $\mathbb{F}$ be a finite field
whose characteristic divides $t$. Let
\[
C^{\left[0\right]}:=C\cap\mathcal{B}\cap A_{\left[e_{n}\right]}\cap\left[\underset{e_{S_{j}}\in\mathcal{B}'}{\bigcap}\left(\underset{i\notin S_{j}}{\sum}A_{e_{i}}+B_{e_{S_{j}}}\right)\right].
\]
We apply to $C^{\left[0\right]}$ the same argument applied to space
$C^{\left\langle 0\right\rangle }$ in the proof of the previous inequality,
we therefore obtain a subspace $C^{\left\{ 0\right\} }:=C^{\left(\left|\mathcal{B}'\right|\right)}$.
Recursively, for $i\in\left[n\right]$, we denote by $C^{\left\{ i\right\} }$,
a subspace of $C^{\left\{ i-1\right\} }$ which is a complementary
space to 
\[
\left(\underset{e_{S_{j}}\in\mathcal{B}',j\neq i}{\sum}B_{e_{S_{j}}}\right)+\left(\underset{e_{j}\in\mathcal{B}'',j\neq i}{\sum}A_{e_{j}}\right)
\]
in
\[
C^{\left\{ i-1\right\} }+\left(\underset{e_{S_{j}}\in\mathcal{B}',j\neq i}{\sum}B_{e_{S_{j}}}\right)+\left(\underset{e_{j}\in\mathcal{B}'',j\neq i}{\sum}A_{e_{j}}\right);
\]
we have
\[
\mathrm{H}\left(C^{\left\{ i-1\right\} }\mid C^{\left\{ i\right\} }\right)\leq\mathrm{I}\left(C;A_{e_{j}},B_{e_{S_{i}}}:e_{S_{j}}\in\mathcal{B}',e_{j}\in\mathcal{B}'',j\neq i\right)
\]
We define $\tilde{C}:=C^{\left\{ n\right\} }$ and the following inequality
is true
\[
\mathrm{H}\left(C\mid\tilde{C}\right)=\mathrm{H}\left(C\mid C^{\left\{ 0\right\} }\right)+\mathrm{H}\left(C^{\left\{ 0\right\} }\mid\tilde{C}\right)
\]
\[
\leq\mathrm{H}\left(C\mid A_{e_{i}},B_{e_{S_{j}}}:e_{i}\in\mathcal{B}'',e_{S_{j}}\in\mathcal{B}'\right)+\mathrm{H}\left(C\mid A_{\left[e_{n}\right]}\right)+\underset{e_{S_{j}}\in\mathcal{B}'}{\sum}\mathrm{H}\left(C\mid A_{e_{i}},B_{e_{S_{j}}}:i\notin S_{j}\right)
\]
\[
+\underset{i}{\sum}\mathrm{I}\left(C;A_{\left[e_{n}\right]-e_{i}}\right)+\underset{e_{S_{h}}\in\mathcal{B}',i\notin S_{h}.}{\sum}\mathrm{I}\left(C;A_{e_{j}},B_{e_{S_{h}}}:j\notin S_{h},j\neq i\right)
\]
\begin{equation}
+\underset{e_{S_{j}}\in\mathcal{B}'}{\sum}\mathrm{I}\left(C;A_{e_{i}},B_{e_{S_{j}}}:i\in S_{j}\right)+\underset{i}{\sum}\mathrm{I}\left(C;A_{e_{j}},B_{e_{S_{i}}}:e_{S_{j}}\in\mathcal{B}',e_{j}\in\mathcal{B}'',j\neq i\right).\label{eq: bound sobre C rayona}
\end{equation}
We remark that vector subspaces $A_{e_{1}},\ldots,A_{e_{n}}$, $B_{e_{S_{j_{1}}}},\ldots,B_{e_{S_{j_{\left|\mathcal{B}'\right|}}}}$
and $\tilde{C}$ satisfy hypothesis in Lemma \ref{lema desigualdad con kerner}.
Thus,
\begin{equation}
\left[1+\underset{i}{\sum}\left|\mathcal{B}_{e_{i}}\right|\right]\mathrm{H}\left(\tilde{C}\right)\leq\underset{i}{\sum}\left|\mathcal{B}_{e_{i}}\right|\mathrm{H}\left(A_{e_{i}}\right)+\mathrm{H}\left(\mathrm{ker}\left(\varphi_{B}\right)\right).\label{eq: dependiente}
\end{equation}
As $B$ is singular over fields whose characteristic divides $t$,
without loss generality, we suppose that there exists a submatrix
$B_{X}$ such that $e_{S_{k}}$ is a linear combination of the columns
of $B_{X}$. So, from Corollary \ref{Capitulo Tecnica 2 Corolario  importante que usa estructura de acceso de divisores de t},
\begin{equation}
\mathrm{H}\left(\mathrm{ker}\left(\varphi_{B}\right)\right)\leq\mathrm{I}\left(B_{e_{S_{k}}};A_{e_{i}},B_{e_{S_{j}}}:e_{i}\in B_{X}\cap\mathcal{B}'',e_{S_{j}}\in B_{X}\cap\mathcal{B}'\right).\label{eq: desigualdad esk}
\end{equation}
We note that $\tilde{C}\leq\mathcal{B}$ and $\tilde{C}\cap\mathcal{B}_{Y}=O$,
for all $B_{Y}$, where $\mathcal{B}_{Y}$ is the sum of all vector
subspaces indexed by the columns of a proper submatrix $B_{Y}$ of
$B$. From Lemma \ref{Capitulo Tecnica 2 Lema elemental para derivar de una estructuras de acceso  unos espacios interesantes},
taking $C:=\tilde{C}$, $A_{i}:=A_{e_{i}}$, $A_{j}:=B_{e_{S_{j}}}$
according to $e_{i}\in\mathcal{B}''$ or $e_{S_{j}}\in\mathcal{B}'$,
the inequality
\[
\mathrm{H}\left(\tilde{C}\right)+\mathrm{I}\left(B_{e_{S_{k}}};A_{e_{i}},B_{e_{S_{j}}}:e_{i}\in\mathcal{B}'',e_{S_{j}}\in\mathcal{B}',S_{j}\neq S_{k}\right)\leq\mathrm{H}\left(B_{e_{S_{k}}}\right)
\]
\[
\text{\,\,\,\,\,\,\,\,\,\,\,\,\,\,\,\,\,\,\,\,\,\,\,\,\,\,\,\,\,\,\,\,\,\,\,\,\,\,\,\,\,\,\,\,\,\,\,\,\,\,\,\,\,\,\,\,\,\,\,\,\,\,\,\,\,\,\,\,\,\,\,\,\,\,\,\,\,\, [we note that \ensuremath{\bar{B}_{e_{S_{k}}}\cap\mathcal{B}_{B-k}=O} and \ensuremath{\mathrm{H}\left(\tilde{C}\right)=\mathrm{H}\left(\bar{B}_{e_{S_{k}}}\right)}]}
\]
which implies
\[
\mathrm{H}\left(\tilde{C}\right)+\mathrm{I}\left(B_{e_{S_{k}}};A_{e_{i}},B_{e_{S_{i}}}:e_{i}\in B_{X}\cap\mathcal{B}'',e_{S_{i}}\in B_{X}\cap\mathcal{B}'\right)\leq\mathrm{H}\left(B_{e_{S_{k}}}\right).
\]
Using inequality (\ref{eq: desigualdad esk}), we have
\[
\mathrm{H}\left(\tilde{C}\right)+\mathrm{H}\left(\mathrm{ker}\left(\varphi_{B}\right)\right)\leq\mathrm{H}\left(B_{e_{S_{k}}}\right).
\]
Therefore, from inequality (\ref{eq: dependiente}),
\[
\left[2+\underset{i}{\sum}\left|\mathcal{B}_{e_{i}}\right|\right]\mathrm{H}\left(\tilde{C}\right)\leq\underset{i}{\sum}\left|\mathcal{B}_{e_{i}}\right|\mathrm{H}\left(A_{e_{i}}\right)+\mathrm{H}\left(B_{e_{S_{k}}}\right).
\]
From this and inequality (\ref{eq: bound sobre C rayona}), we obtain
the desired inequality:
\[
\mathrm{H}\left(C\right)-\mathrm{H}\left(C\mid A_{\left[e_{n}\right]}\right)-\underset{e_{S_{j}}\in\mathcal{B}'}{\sum}\mathrm{H}\left(C\mid A_{e_{i}},B_{e_{S_{j}}}:i\notin S_{j}\right)-\mathrm{H}\left(C\mid A_{e_{i}},B_{e_{S_{j}}}:e_{i}\in\mathcal{B}'',e_{S_{j}}\in\mathcal{B}'\right)
\]
\[
-\underset{i}{\sum}\mathrm{I}\left(C;A_{\left[e_{n}\right]-e_{i}}\right)-\underset{e_{S_{h}}\in\mathcal{B}',i\notin S_{h}.}{\sum}\mathrm{I}\left(C;A_{e_{j}},B_{e_{S_{h}}}:j\notin S_{h},j\neq i\right)
\]
\[
-\underset{e_{S_{j}}\in\mathcal{B}'}{\sum}\mathrm{I}\left(C;A_{e_{i}},B_{e_{S_{j}}}:i\in S_{j}\right)-\underset{i}{\sum}\mathrm{I}\left(C;A_{e_{i}},B_{e_{S_{j}}}:e_{S_{j}}\in\mathcal{B}',e_{i}\in\mathcal{B}'',j\neq i\right)
\]
\[
\leq\mathrm{H}\left(\tilde{C}\right)\leq\frac{1}{2+\underset{i}{\sum}\left|\mathcal{B}_{e_{i}}\right|}\left(\underset{i}{\sum}\left|\mathcal{B}_{e_{i}}\right|\mathrm{H}\left(A_{e_{i}}\right)+\mathrm{H}\left(B_{e_{S_{k}}}\right)\right).
\]
\end{proof}

\section{Examples and applications}

\begin{figure}[h]
\[
\begin{array}{c}
B_{1}\cdots B_{t+1}A_{t+2}\cdots A_{M\left(n,t\right)}\\
\left(\begin{array}{cccccc}
0 & \cdots & 1 & 0 & \cdots & 0\\
1 & \vdots & 1 & 0 & \vdots & 0\\
\vdots & \vdots & \vdots & \vdots & \vdots & 0\\
1 & \vdots & 1 & 0 & \vdots & 0\\
1 & \vdots & 0 & 0 & \vdots & \vdots\\
1 & \vdots & 1 & 1 & \vdots & 0\\
1 & \vdots & \vdots & 0 & \vdots & 0\\
1 & \vdots & 1 & \vdots & \vdots & 0\\
1 & \cdots & 1 & 0 & \cdots & 1
\end{array}\right)
\end{array}\,\,\,\,\,\,\,\,\,\,\,\,\,\,\,\,\,\,\begin{array}{c}
a_{1}\,\,\cdots\,\,a_{t+1}\,\,b_{1}\,\,\cdots\,\,b_{t+1}\,\,c\\
\left(\begin{array}{ccccccc}
1 & \cdots & 0 & 0 & \cdots & 1 & 1\\
0 & \vdots & \vdots & 1 & \vdots & 1 & \vdots\\
\vdots & \vdots & \vdots & \vdots & \vdots & \vdots & \vdots\\
0 & \vdots & 0 & 1 & \vdots & 1 & \vdots\\
0 & \cdots & 1 & 1 & \cdots & 0 & 1
\end{array}\right)
\end{array}
\]

\caption{\label{fig:Matriz para la aplicaci=0000F3n de la tecnica 2}A family
of matrices $B_{M\left(n,t\right)}^{t}$ whose determinant is $\pm t$
and \label{fig:grafica matriz de representaci=0000F3n para la aplicaci=0000F3n en secret sharing}
a family of representable matroids.}
\end{figure}

We now produce some characteristic-dependent linear rank inequalities
using a convenient class of matrices. Let $n\geq7$ and $t$ integer
such that $2\leq t\leq\left\lfloor \frac{n-1}{2}\right\rfloor -1$
and $M\left(n,t\right)=n-t-2$. In Theorem \ref{thm:Teorema central del metodo 2},
we take square matrices $B_{M\left(n,t\right)}^{t}$ as described
in figure \ref{fig:Matriz para la aplicaci=0000F3n de la tecnica 2}
on the left side with column vectors of the form $A_{i}:=A_{e_{i}}=e_{i}$,
$B_{i}:=B_{e_{\left[M\left(n,t\right)\right]-i}}=c-e_{i}$ and $c=\underset{i\in\left[M\left(n,t\right)\right]}{\sum}e_{i}$.
We have $\left|\det\left(B_{M\left(n,t\right)}^{t}\right)\right|=t$,
$\left|\mathcal{B}_{B_{M\left(n,t\right)}^{t}}'\right|=t+1$, $\left|\mathcal{B}_{B_{M\left(n,t\right)}^{t}}''\right|=M\left(n,t\right)-t-1$,
$\left|\mathcal{B}_{e_{i}}\right|=1$ for $i\in\left[t+1\right]$
and $\left|\mathcal{B}_{e_{i}}\right|=0$ for $i\in\left[t+2,M\left(n,t\right)\right]$.
Let $A_{1}$, $A_{2}$, $\ldots$, $A_{M\left(n,t\right)}$, $B_{1}$,
$B_{2}$, $\ldots$, $B_{t+1}$ and $C$ be vector subspaces of a
finite dimensional vector space $V$ over a finite field $\mathbb{F}$.
We have

- A characteristic-dependent linear rank inequality over fields whose
characteristic does not divide $t$,
\[
\mathrm{H}\left(C\right)\leq\frac{1}{t+2}\underset{i\in\left[t+1\right]}{\sum}\mathrm{H}\left(A_{i}\right)+\mathrm{I}\left(C;B_{\left[t+1\right]},A_{\left[t+2,M\left(n,t\right)\right]}\right)+\mathrm{H}\left(C\mid A_{\left[M\left(n,t\right)\right]}\right)
\]
\[
+\underset{i\in\left[M\left(n,t\right)\right]}{\sum}\mathrm{I}\left(C;A_{\left[M\left(n,t\right)\right]-i}\right)+\underset{i\in\left[t+1\right]}{\sum}\left[\mathrm{I}\left(C;B_{i}\right)+\mathrm{H}\left(C\mid A_{i},B_{i}\right)+\mathrm{I}\left(C;A_{\left[M\left(n,t\right)\right]-i},B_{i}\right)\right].
\]

- A characteristic-dependent linear rank inequality over fields whose
characteristic divides $t$,
\[
\mathrm{H}\left(C\right)\leq\frac{1}{t+3}\left[\underset{i\in\left[t+1\right]}{\sum}\mathrm{H}\left(A_{i}\right)+\mathrm{H}\left(B_{1}\right)\right]+\mathrm{H}\left(C\mid B_{\left[t+1\right]},A_{\left[t+2,M\left(n,t\right)\right]}\right)+\mathrm{H}\left(C\mid A_{\left[M\left(n,t\right)\right]}\right)
\]
\[
+\underset{i\in\left[M\left(n,t\right)\right]}{\sum}\mathrm{I}\left(C;A_{\left[M\left(n,t\right)\right]-i}\right)+\underset{i\in\left[t+1\right]}{\sum}\left[\mathrm{I}\left(C;B_{i}\right)+\mathrm{H}\left(C\mid A_{i},B_{i}\right)+\mathrm{I}\left(C;A_{\left[M\left(n,t\right)\right]-i},B_{i}\right)\right]
\]
\[
+\underset{i\in\left[t+1\right]}{\sum}\mathrm{I}\left(C;B_{\left[t+1\right]-i},A_{\left[t+2,M\left(n,t\right)\right]}\right)+\underset{i\in\left[t+2,M\left(n,t\right)\right]}{\sum}\mathrm{I}\left(C;B_{\left[t+1\right]},A_{\left[t+2,M\left(n,t\right)\right]-i}\right).
\]

\begin{rem}
We produce a class of $\left\lfloor \frac{n-1}{2}\right\rfloor -2$
inequalities that are true over finite sets of primes and another
class of $\left\lfloor \frac{n-1}{2}\right\rfloor -2$ inequalities
that are true over co-finite sets of primes.
\end{rem}
Let $t\in\mathbb{N}$, $t>1$ and let $\mathbb{F}$ be a finite field.
We use the port at $c$ of the representable matroid obtained from
the matrix in Figure \ref{fig:grafica matriz de representaci=0000F3n para la aplicaci=0000F3n en secret sharing}
on the right side; the set of participants is $P=\left\{ a_{1},\ldots,a_{t+1},b_{1},\ldots,b_{t+1}\right\} $,
with dealer $p=c$.

When $\mathrm{char}\left(\mathbb{F}\right)$ divides $t$, the following
set is a subclass of the minimal qualified set:
\[
\left\{ a_{1}b_{1},\ldots,a_{t+1}b_{t+1},a_{1}\cdots a_{t+1},a_{1}b_{2}\cdots b_{t+1},b_{1}a_{2}b_{3}\cdots b_{t+1},\ldots,b_{1}\cdots b_{t}a_{t+1}\right\} ,
\]
and the following set is a subclass of the non-qualified set:
\[
\left\{ b_{1}a_{2}\cdots a_{t+1},a_{1}b_{2}a_{3}\cdots a_{t+1},\ldots,a_{1}\cdots a_{t}b_{t+1}\right\} \cup\left\{ b_{1}\cdots b_{t+1}\right\} .
\]
When $\mathrm{char}\left(\mathbb{F}\right)$ does not divide $t$,
the following set is a subclass of the minimal qualified set:
\[
\left\{ a_{1}b_{1},\ldots,a_{t+1}b_{t+1},a_{1}\cdots a_{t+1},a_{1}b_{2}\cdots b_{t+1},b_{1}a_{2}b_{3}\cdots b_{t+1},\ldots,b_{1}\cdots b_{t}a_{t+1}\right\} \cup\left\{ b_{1}\cdots b_{t+1}\right\} ,
\]
and the following set is a subclass of the non-qualified set:
\[
\left\{ b_{1}a_{2}\cdots a_{t+1},a_{1}b_{2}a_{3}\cdots a_{t+1},\ldots,a_{1}\cdots a_{t}b_{t+1}\right\} .
\]
So, we have defined two types of access structures using matroid ports.
Let $\mathcal{F}_{t}$ be an access structure of the first type and
let $\mathcal{N}_{t}$ be an access structure of the second type.
We remark that $\mathcal{F}_{2}$ is a port of Fano matroid and $\mathcal{N}_{2}$
is a port of non-Fano matroid. Some characteristic-dependent linear
rank inequalities are used for getting lower bounds on the linear
information ratio over specific fields of these access structures.
Taking $n=2t+3$ and $M\left(n,t\right)=t+1$ in previous inequalities,
we obtain two classes of constraints. The first must be satisfied
by linear secret sharing schemes over fields whose characteristic
does not divide $t$ and the second must be satisfied by linear secret
sharing schemes over fields whose characteristic divides $t$. We
have the following proposition.
\begin{prop}
Let $t\in\mathbb{N}$, $t>1$ and let $\mathbb{F}$ be a finite field.
For any $\mathcal{F}_{t}$ and $\mathcal{N}_{t}$, we have:
\begin{itemize}
\item $\sigma\left(\mathcal{F}_{t}\right)=\lambda_{\mathrm{char}\left(\mathbb{F}\right)\mid t}\left(\mathcal{F}_{t}\right)=\kappa\left(\mathcal{F}_{t}\right)=\kappa_{\mathrm{char}\left(\mathbb{F}\right)\mid t}^{*}\left(\mathcal{F}_{t}\right)=1$.
\item $\sigma\left(\mathcal{N}_{t}\right)=\lambda_{\mathrm{char}\left(\mathbb{F}\right)\nmid t}\left(\mathcal{N}_{t}\right)=\kappa\left(\mathcal{N}_{t}\right)=\kappa_{\mathrm{char}\left(\mathbb{F}\right)\nmid t}^{*}\left(\mathcal{N}_{t}\right)=1$.
\item $\lambda_{\mathrm{char}\left(\mathbb{F}\right)\nmid t}\left(\mathcal{F}_{t}\right)\geq\kappa_{\mathrm{char}\left(\mathbb{F}\right)\nmid t}^{*}\left(\mathcal{F}_{t}\right)\geq\frac{t+2}{t+1}$.
\item $\lambda_{\mathrm{char}\left(\mathbb{F}\right)\mid t}\left(\mathcal{N}_{t}\right)\geq\kappa_{\mathrm{char}\left(\mathbb{F}\right)\mid t}^{*}\left(\mathcal{N}_{t}\right)\geq\frac{t+3}{t+2}.$
\end{itemize}
\end{prop}
\begin{proof}
It is clear that these access structures are ideal over fields where
the associated matroid are representable. So, we have the optimal
information ratios are equal to $1$ over these fields. It remains
for proving the last two items. Taking $A_{i}=a_{i}$, $B_{i}=b_{i}$
and $C=c$ in the linear programming problem \ref{prob:problema de programaci=0000F3n lineal para secret sharing}
with the constraints valid over fields whose characteristic does not
divide $t$, The access structure $\mathcal{F}_{t}$ holds that $f\left(a_{i}\right)\leq v$,
$f\left(a_{i}\right)\leq v$, $f\left(\emptyset\right)=0$, $f\left(c\right)=1$,
$f\left(c\mid a_{\left[t+1\right]}\right)=f\left(c\mid a_{i},b_{i}\right)=f\left(c;a_{\left[M\left(n,t\right)\right]-i},b_{i}\right)$
$=f\left(c;a_{\left[t+1\right]-i}\right)=f\left(c\mid a_{i},b_{i}\right)=f\left(c;a_{\left[t+1\right]-i},b_{i}\right)=0$
. Thus, using the constraint obtained from the characteristic-dependent
linear rank inequality $n=2t+3$ and $M\left(n,t\right)=t+1$, we
get
\[
1=f\left(c\right)\leq\frac{1}{t+2}\underset{i\in\left[t+1\right]}{\sum}f\left(a_{i}\right)\leq\frac{t+1}{t+2}v.
\]
Therefore, $\kappa_{\mathrm{char}\left(\mathbb{F}\right)\nmid t}^{*}\left(\mathcal{F}_{t}\right)\geq v\geq\frac{t+2}{t+1}$.
In a similar way, we show the other inequalities.
\end{proof}

\section*{Acknowledgments}

The author thanks the support provided by COLCIENCIAS in Conv. 727
and Carles Padró for the doctoral stay in Barcelona.

\end{document}